\documentclass[aps,twocolumn,showpacs]{revtex4-1}
\usepackage{amsmath}

\usepackage{color}
\usepackage{graphicx}

\begin{document}

\bibliographystyle{utcaps}
\title{Stable Quantum Monte Carlo Simulations for  Entanglement Spectra of Interacting Fermions }
\author{\firstname{Fakher F.} \surname{Assaad}}
\affiliation{Institut f\"ur Theoretische Physik und Astrophysik, Universit\"at W\"urzburg, Am Hubland, D-97074 W\"urzburg, Germany}
\date{\today}

\begin{abstract}
We show that the two recently proposed methods to compute  Renyi  entanglement entropies in the realm of    determinant quantum Monte Carlo   methods for fermions are  in principle equivalent, but differ in sampling strategies.   The analogy allows to formulate  a numerically stable  calculation of the entanglement spectrum at strong coupling.  We demonstrate  the approach by studying static and dynamical properties of the entanglement hamiltonian across the   interaction  driven  quantum phase transition  between  a topological insulator and quantum  antiferromagnet  in the Kane-Mele Hubbard model.  The formulation is not limited to fermion systems  and can readily  be  adapted to world-line based simulations of  bosonic systems. 
\end{abstract}


\maketitle

\section{Introduction}

Consider a bipartition of a Hilbert space of a  many body system  in a state  described by a density matrix $\hat{\rho}$.  Tracing over   the degrees of freedom  of one partition  defines a reduced density matrix.   Its entropy  provides a measure of the entanglement between the two partitions \cite{Amico08}.  At zero  temperature one generically expects  the entanglement entropy to follow an area law \cite{Eisert10}. Corrections to this law  have the potential of revealing fundamental  properties  such as topological order  \cite{Levin06,Kitaev06_1,Isakov11}   or  the central charge for one-dimensional systems \cite{Calabrese04}.   The logarithm of the reduced density matrix defines  an entanglement  Hamiltonian \cite{Haldane08}, the study of which has spurred  substantial research  \cite{Lauchli10,Thomale10,Qi12,Fidkowski10,Turner10,Kolley13}.  The notion that it contains fundamental  and universal information has emerged  and  has been critically discussed  \cite{Chandran13}.   The aim of this article is to develop tools  to study the properties on the entanglement Hamiltonian  in the realm of quantum Monte Carlo (QMC)  simulations for fermions.  

For fermonic systems the calculation of the Renyi entanglement entropy has followed two different routes.  One  method builds  on a replica idea   with sampling based on a {\it swap}  move \cite{Hastings10,Humeniuk12}.  This approach was initially proposed for spin systems \cite{Hastings10,Humeniuk12}   at zero and finite temperatures and then  generalized to fermions in the realm of determinant   \cite{Broecker14} and continuous time \cite{Wang14} quantum Monte Carlo (QMC)  methods. We will refer to this  algorithm as the swap algorithm.   The  other  approach put forward in Ref.  \cite{Grover13}   utilizes the fact that in auxiliary field algorithms \cite{Blankenbecler81} -- which express the interacting system in terms of a sum of non-interacting problems,  the  density matrix can be formally written as a sum over gaussian operators  \cite{Peschel03}.  We will refer to this algorithm as the gaussian  approach.  It is  in principle simple to implement and allows for generalizations to compute entanglement spectra \cite{Assaad13a}.   As  pointed out in \cite{Assaad13a,Broecker14}   it suffers from an exponential growth of fluctuations  in the strong coupling limit and  when the subsystem size is {\it large}.   

We will show that within the auxiliary field  approach  both methods are equivalent, and  merely correspond to  different ways of carrying out the sampling. Since the swap algorithm is  more stable than  the gaussian  one,  the equivalence of the two methods shows how to stabilize the gaussian algorithm.   As a consequence  we are able to formulate a stable  QMC algorithm  allowing a detailed study of  the entanglement Hamiltonian  for fermion systems at strong coupling.

	 Here we will demonstrate  the  validity of the approach   by studying a previously not accessible parameter region of the Kane-Mele Hubbard model \cite{KaneMele05,KaneMele05b,HoAsreview2013,Hohenadler10,Hohenadler12,Assaad12}.  In particular, we will  concentrate on   the correlation induced  phase transition from a topological insulator to a quantum antiferromagnetic   from the perspective of the entanglement spectrum both  in the single particle and particle-hole sectors. 
	 
	 The article is organized as follows. In the next section we will show the equivalence of  the swap and gaussian algorithms.   Section~\ref{Entanglemet.Sec} will use this  equivalence to reformulate the proposed evaluation of the entanglement spectra of Ref.~\cite{Assaad13a} in a numerically stable manner. Before concluding in Sec.~\ref{Conc.Sec}, we test our approach   by studying  the correlation driven quantum phase transition in the Kane-Mele Hubbard model from the perspective of the entanglement spectrum.

\section{Equivalence  swap and gaussian algorithms for the $n^{th}$ Renyi entropy.}
Here we will start with  the swap algorithm formulation of the $n^{th}$ Renyi entropy and derive the gaussian  algorithm of Ref. \cite{Grover13}. 
We consider a real space partitioning of the Hilbert space,  $\cal{ H}  = \cal{ H}_A \otimes \cal{H}_B$. To compute the  n$^\text{th}$ Renyi entropy, 
\begin{equation}
	S_n = -\frac{1}{n-1}  \ln  \text {Tr}_{{\cal H}_A}  \hat{\rho}_{A}^n, 
\end{equation}
with $ \hat{\rho}_{A}  = {\text Tr}_{{\cal H}_B}   \hat{\rho} $ and $ \hat{\rho}$ the density matrix, we consider  the replicated  Hilbert  space:
\begin{equation}
	{\cal H}_{ \text{tot}} = {\cal H}_{A}  \otimes {\cal H}^{(1)}_{B} \otimes {\cal H}^{(2)}_{B}  \cdots {\cal H}^{(n)}_{B}. 
\end{equation}
At $n=1$,  ${\cal H}_{ \text{tot}} $  reduces to the original Hilbert  space, ${\cal H} =  {\cal H}_{A}  \otimes {\cal H}_{B} $, and the Hamiltonian we will consider reads 
\begin{equation}
	\hat{H}  =   \sum_{\alpha}  \hat{h}^{(\alpha)}_{A} \otimes \hat{h}^{(\alpha)}_{B}.  
\end{equation}

\begin{figure}
   \includegraphics[width=0.8\linewidth]{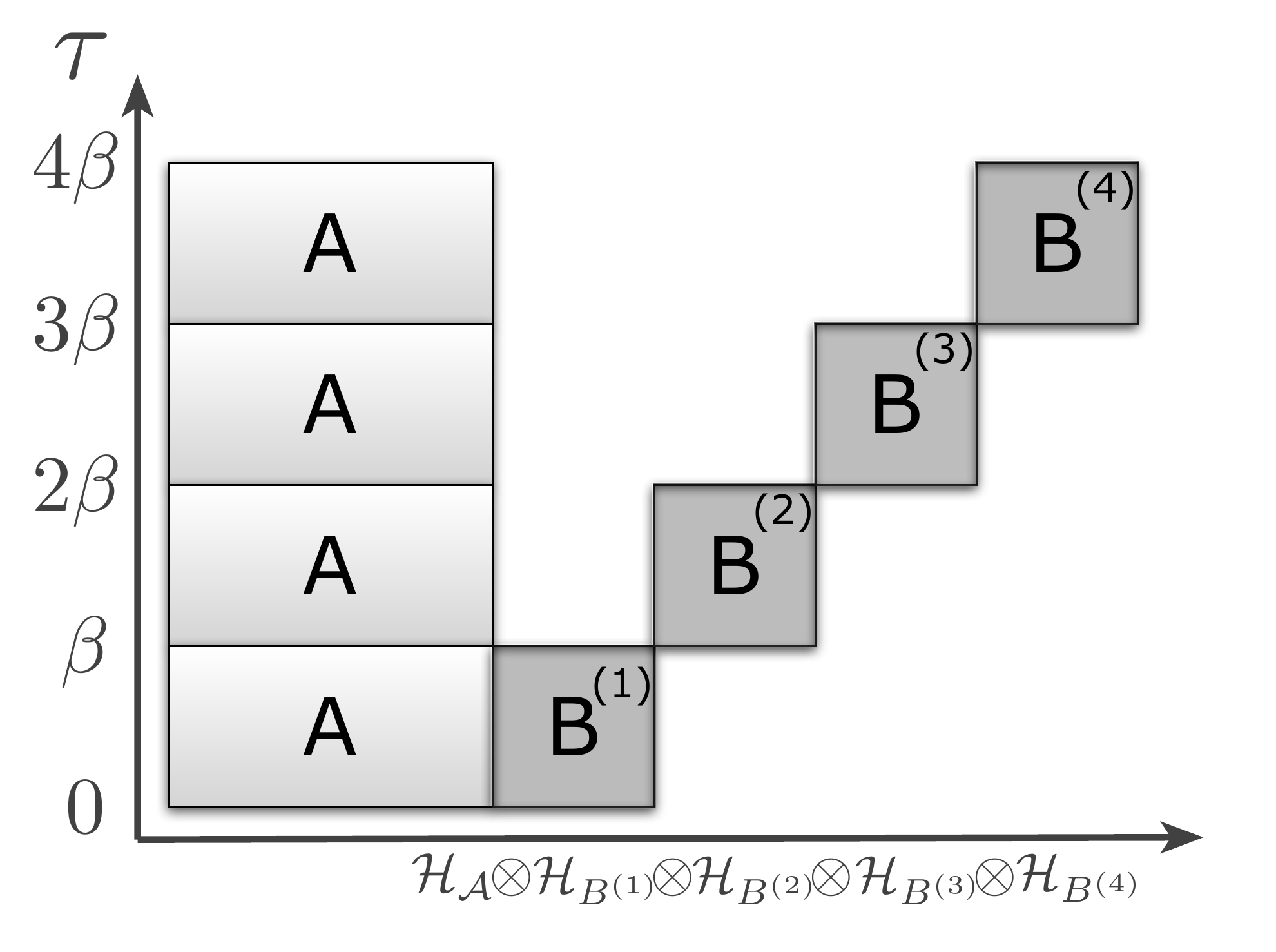} 
   \caption{ Schematic view  of the imaginary time propagation  of the Hamiltonian $\hat{H}(\tau) $ for the case $n=4$. The  Hamiltonian vanishes in the non-shaded regions. 
   \label{Htau.fig}}
\end{figure}

In the swap algorithm one expands the imaginary time propagation from $\beta$ to $n \beta$ ( i.e. $ \tau \in  \left[ 0, n \beta \right] $) and defines a time dependent Hamiltonian in the Hilbert space $\cal{H}_{\text{tot}} $  as:
\begin{equation}
	 \hat{H}(\tau)   =    \sum_{r=1}^{n}   \Theta \left[  \tau - (r-1) \beta \right] \Theta \left[  r \beta - \tau \right]   \hat{H}^{(r)}.
\end{equation}
Here   
\begin{eqnarray}
	\hat{H}^{(r)}& &    =   \\
	   & &   \sum_{\alpha} \hat{h}^{(\alpha)}_{A} \otimes  \hat{1}^{(1)}_B \cdots   \otimes   \hat{1}^{(r-1)}_B \otimes  \hat{h}^{(\alpha)}_{B}  \otimes  \hat{1}^{(r+1)}_B \cdots  \otimes  \hat{1}^{(n)}_B.  \nonumber 
\end{eqnarray}
A schematic representation of this time evolution is given in Fig.~\ref{Htau.fig}. 
With this construction,   one will show that: 
\begin{eqnarray}
\label{Rhon.eq}
	\text {Tr}_{{\cal H}_{A}}  \hat{\rho}_{A}^n &  = &    \frac{1}{Z^n}  {\text Tr}_{ { \cal H}_{tot} } {\cal T} e^{-\int_{0}^{ n \beta}  d \tau \hat{H}(\tau)}   \nonumber \\
       &   \equiv &  \frac{1}{Z^n}  {\text Tr}_{ { \cal H}_{tot} } e^{-\beta \hat{H}^{(n)}} \cdots e^{-\beta \hat{H}^{(1)} }.
\end{eqnarray}
The above follows from writing the trace
\begin{eqnarray}
    &  & { \text  Tr}_{H_{tot}}   \left[  \hat{O} \right]  =   \\ 
    &  & \sum_{A,B^{(1)}, \cdots ,B^{(n)}}  \langle A,B^{(1)}, \cdots ,B^{(n) } | \hat{O} |  A,B^{(1)}, \cdots ,B^{(n) }  \rangle \nonumber
 \end{eqnarray}
 where $ A $ and  $B$ run over a complete set of orthonormal states of ${\cal H}_{A}$ and ${\cal H}_B$ respectively and by noting that: 
 \begin{eqnarray}
 	& &  \langle A,B^{(1)}, \cdots ,B^{(n) } | e^{-\beta \hat{H}^{(r)}} |  A_1,B_1^{(1)}, \cdots ,B_1^{(n) }   \rangle =   \nonumber   \\
	& &  \langle A,B^{(r)} | e^{-\beta \hat{h}^{(r)}} |  A_1,B_1^{(r)}   \rangle   \prod_{\stackrel{i = 1 }{i \ne r }}^{n}  \delta_{B^{(i)},B_1^{(i)}}.
 \end{eqnarray}
Here  
$\hat{H}^{(r)} \equiv   \hat{h}^{(r)} \bigotimes_{\stackrel{i = 1 }{i \ne r }}^{n}  \hat{1}_B^{(i)} $ such that  $ \hat{h}^{(r)} $  corresponds to the Hamiltonian in Hilbert space 
${\cal H}_A \otimes {\cal H}_{B}^{(r)}$.  One can now explicitly compute the trace in Eq.~(\ref{Rhon.eq}) by inserting a  complete set of states in ${\cal H}_{tot}$ between each replica so as to obtain:  
\begin{widetext}
\begin{eqnarray}
 {\text Tr}_{ { \cal H}_{tot} } & &  e^{-\beta \hat{H}^{(n)}} \cdots e^{-\beta \hat{H}^{(1)} } = \sum_{A_1 \cdots A_n}  
 \sum_{B_1 \cdots B_n}   \langle  A_1,B^{(n)}_n | e^{-\beta \hat{h}^{(n)}} |  A_n,B_n^{(n)}  \rangle     
                                                \langle A_n,B^{(n-1)}_{n-1} | e^{-\beta \hat{h}^{(n-1)}} |  A_{n-1},B_{n-1}^{(n-1)}  \rangle  \cdots   \times \nonumber  \\
                                             & &       \langle A_2,B^{(1)}_{1} | e^{-\beta \hat{h}^{(1)}} |  A_{1},B_1^{(1)} \rangle  = Z^{n} \sum_{A_1 \cdots A_n}   
                          \langle  A_1 | \hat{\rho}_A  |  A_n \rangle     \langle  A_n | \hat{\rho}_A  |  A_{n-1} \rangle    \cdots  \langle  A_2 | \hat{\rho}_A  |  A_{1} \rangle 
                          = Z^{n}  {\text Tr}_{{ \cal H}_{A} } \hat{\rho}_A^n 
\end{eqnarray}
\end{widetext}
Note that the reduced  density matrix $  \langle  A  |  \hat{\rho}_A | A' \rangle = \frac{1}{Z}\sum_{B^{(r)}} \langle A ,B^{(r)} | e^{-\beta \hat{h}^{(r)}} |  A',B^{(r)} \rangle  $ 
is independent on the choice of the replica.  
 The partition function $Z$ of the original Hamiltonian can be written  as:
 \begin{equation}
 	Z  =   {\text Tr}_{{\cal H}_{tot}} \left[ e^{- \beta \hat{H}^{(r)}} \right]  d^{-(n-1)N_B}.
 \end{equation}
 $d$ corresponds to the number of states per site ( $d=4$ for the spin-$1/2$ Hubbard model)  and $N_B$ the number of sites in the partition $B$ such that 
 $d^{(n-1)N_B} $  counts the number of states in the Hilbert space  ${\cal H} =  \bigotimes_{r=1}^{n-1}  {\cal H }_B^{(r)}$ .  Hence,  the  factor $d^{-(n-1)N_B}$ compensates the over  counting   when  computing the partition function of the original Hamiltonian by tracing over $ {{\cal H}_{tot}} $.
Note again that the partition function does not depend on the specific choice of the replica. \\
Thus, 
\begin{eqnarray}
	{\text  Tr}_{{\cal H}_{A}} & &   \hat{\rho}_{A}^n =  \\ 
		& & \frac{ {\text  Tr}_{{\cal H}_{tot}} \left[  e^{-\beta \hat{H}^{(n)}} \cdots e^{-\beta \hat{H}^{(1)}} \right] }
	     {   {\text  Tr}_{{\cal H}_{tot}} \left[  e^{-\beta \hat{H}^{(n)}}  \right] \cdots {\text  Tr}_{{\cal H}_{tot}} \left[  e^{-\beta \hat{H}^{(1)}}  \right] d^{-n(n-1)N_B} } \nonumber
\end{eqnarray}

We are now in the position to compute  the  Renyi entropy with  auxiliary field quantum Monte Carlo methods. Here, we will use the finite temperature algorithm  \cite{White89,Assaad08}. For a given replica, we can make use of the Trotter 
decomposition  so as to write 
\begin{equation}
	e^{-\beta \hat{H}^{(r)}} = \prod_{\tau=1}^{L_{\tau}} e^{-\Delta \tau \hat{T}^{(r)}/2}  e^{-\Delta \tau \hat{H}_{U}^{(r)}}   e^{-\Delta \tau \hat{T}^{(r)}/2}    + {\cal O}  \left(  \Delta \tau^2\right)
 \end{equation} 
 and the  Hubbard Stratonovitch transformation 
 \begin{equation}
 e^{-\Delta \tau \hat{H}_{U}^{(r)}} = \sum_{ \pmb{s}_{\tau}^{(r)}}  e^{ \hat{V}^{(r)} ( \pmb{s}_{\tau}^{(r)}) }. 
 \end{equation}
 For each imaginary time and replica, we have a vector of Hubbard Stratonovitch fields, $ \pmb{s}_{\tau}^{(r)} $. It is important to  remember that, by construction,  the  dimension of 
 $ \pmb{s}_{\tau}^{(r)} $  is  identical to that  of a single simulation at   $n=1$. For the Hubbard model,  $ \pmb{s}_{\tau}^{(r)}  $  corresponds to a vector  of length $N_A + N_B$ of Ising spins  and we have used a  transformation where the Ising field couples to the local density \cite{Hirsch83}.
$ \hat{T}^{(r)}  $ and  $\hat{V}^{(r)}( \pmb{s}_{\tau}^{(r)}) $  are single particle operators  which one can write as: 
\begin{equation}
   \hat{T}^{(r)}    = \hat{{\pmb c}}^{\dagger}  T^{(r)} \hat{{\pmb c}}   \text{     and    } 
   \hat{V}^{(r)} (  \pmb{s}_{\tau}^{(r)}  )    = \hat{{\pmb c}}^{\dagger}   V^{(r)} (  \pmb{s}_{\tau}^{(r)}  )   \hat{{\pmb c}}.
\end{equation}
Here, $ \hat{{\pmb c}}$ is a vector of  fermionic annihilation operators   running over all single particle states of the Hilbert space  ${\cal H}_{tot}$.
The imaginary time propagation now reads:
 \begin{eqnarray}
 	& & e^{-\beta \hat{H}^{(r)}}  =  \\ 
	& & \sum_{\pmb{s}^{(r)}}   \prod_{\tau=1}^{L_{\tau}}  e^{-\Delta \tau \hat{T}^{(r)}/2}   e^{ \hat{V}^{(r)} ( \pmb{s}_{\tau}^{(r)}) }   e^{-\Delta \tau \hat{T}^{(r)}/2} \equiv 
	   \sum_{\pmb{s}^{(r)}}   \hat{U}^{(r)}_{\pmb{s}^{(r)}} \nonumber 
 \end{eqnarray}
where $\pmb{s}^{(r)} $  is a short hand notation for $ \pmb{s}^{(r)}_{1} \cdots \pmb{s}^{(r)}_{L_\tau}$. 

With the above, we can compute   the Renyi entropy as: 
\begin{widetext}
\begin{equation}
\label{Swap.eq}
{\text  Tr}_{{\cal H}_{A}}  \hat{\rho}_{A}^n = \frac{ \sum_{\pmb{s}^{(1)} \cdots \pmb{s}^{(n)} }  {\text Tr}_{{\cal H}_{tot}} \left[ \hat{U}^{(n)}_{\pmb{s}^{(n)}} \cdots \hat{U}^{(1)}_{\pmb{s}^{(1)}} \right]}  { \sum_{\pmb{s}^{(1)} \cdots \pmb{s}^{(n)}}   {\text Tr}_{{\cal H}_{tot}} \left[ \hat{U}^{(n)}_{\pmb{s}^{(n)}} \right] \cdots \cdots {\text Tr}_{{\cal H}_{tot}} \left[ \hat{U}^{(1)}_{\pmb{s}^{(1)}} \right]  
 d^{-n(n-1)N_B} }  
\end{equation}
\end{widetext}
$ {\text  Tr}_{{\cal H}_{A}}  \hat{\rho}_{A}^n $ corresponds to the ratio of two { \it partition functions},   defined  on the same configuration space. 
Note  that the symmetries  which ensure  the absence of sign problem for the original Hamiltonian  can be used to prove  the absence of sign problem  for the numerator.  For the Kane-Mele Hubbard model we refer the reader to \cite{Hohenadler10,Zheng11,Hohenadler12}  for a proof  of the absence of sign problem at half-band filling.  
The ratio  in Eq.~(\ref{Swap.eq}) can be computed with the swap 
algorithm described in \cite{Humeniuk12}.   This approach used to compute the Renyi entropies corresponds to the one  adopted for bosonic systems and recently generalized to fermions \cite{Broecker14}.    To show the equivalence to the gaussian algorithm proposed in Ref.  \cite{Grover13}  and further developed in Ref. \cite{Assaad13a} to access entanglement spectrum  we can rewrite Eq. (\ref{Swap.eq}) as: 
\begin{widetext}
\begin{equation}
	{\text  Tr}_{{\cal H}_{A}}  \hat{\rho}_{A}^n =
       \sum_{\pmb{s}^{(1)}, \cdots, \pmb{s}^{(n)} }  P(\pmb{s}^{(1)},\cdots,  \pmb{s}^{(n)}) \langle \langle \hat{O}  \rangle \rangle_{\pmb{s}^{(1)}\cdots \pmb{s}^{(n)}}   
\end{equation}
where
\begin{equation}
P(\pmb{s}^{(1)},\cdots, \pmb{s}^{(n)}) =   \frac{{\text Tr}_{{\cal H}_{tot}} \left[ \hat{U}^{(n)}_{\pmb{s}^{(n)}} \right] \cdots \cdots {\text Tr}_{{\cal H}_{tot}} \left[ \hat{U}^{(1)}_{\pmb{s}^{(1)}} \right] }
                                                               { \sum_{\pmb{s}^{(1)} \cdots \pmb{s}^{(n)} } {\text Tr}_{{\cal H}_{tot}} \left[ \hat{U}^{(n)}_{\pmb{s}^{(n)}} \right] \cdots \cdots {\text Tr}_{{\cal H}_{tot}} \left[ \hat{U}^{(1)}_{\pmb{s}^{(1)}} \right]} 
\end{equation}
and 
\begin{equation}
\langle \langle \hat{O}  \rangle \rangle_{\pmb{s}^{(1)}\cdots \pmb{s}^{(n)}} =
 \frac{  {\text Tr}_{{\cal H}_{tot}} \left[ \hat{U}^{(n)}_{\pmb{s}^{(n)}} \cdots \hat{U}^{(1)}_{\pmb{s}^{(1)}} \right]} 
	        {  {\text Tr}_{{\cal H}_{tot}} \left[ \hat{U}^{(n)}_{\pmb{s}^{(n)}} \right] \cdots \cdots {\text Tr}_{{\cal H}_{tot}} \left[ \hat{U}^{(1)}_{\pmb{s}^{(1)}} \right]  d^{-n(n-1)N_B}  } 
\end{equation}
\end{widetext}
The probability distribution $P(\pmb{s}^{(1)},\cdots, \pmb{s}^{(n)})$   is sampled by carrying out $n$-independent simulations of the original Hamiltonian.   Our task is now to show that 
$\langle \langle \hat{O}  \rangle \rangle_{\pmb{s}^{(1)}\cdots \pmb{s}^{(n)}} $ reduces to Grover's form   \cite{Grover13} for the calculation of the   Renyi entropy.  
\subsection{The $n=2$ case.}
At $n=2$ one can follow  a pedestrian path  
and compute   the ratio of  the two   fermionic determinants.    We will sketch the calculation under the assumption  that  $\hat{U}^{(r)}_{\pmb{s}^{(r)}} $ factorizes into spin-up and spin-down components such that we can only concentrate on the orbital degrees of freedom.  Let $P_{A}$  be a $ (N_A + 2 N_B)  \times  (N_A + 2 N_B) $  matrix  with 
\begin{equation}
\left( P_{A}  \right)_{\pmb{i}, \pmb{j}}   =   
\left\{
\begin{array}{cc}
\delta_{\pmb{i}, \pmb{j}}  &   \text{  if }   \pmb{i} \in A \\
0                                     &   \text{  otherwise }   
\end{array}
\right..
 \end{equation}
Here $\pmb{i}$ and $\pmb{j}$ run over all  the single particle Wannier states of the Hilbert space $ {\cal H}_{A} \otimes  {\cal H}_{B} \otimes   {\cal H}_{B'}  $, and $\pmb{i} \in A$ states that 
Wanier state $\pmb{i}$  belongs to  $ {\cal H}_{A}$.  Clearly $P_A$ is a projector, and we will define similar quantities  $P_B$ and $P_{B'}$. Note that  $ P_A, P_B$ and $P_{B'}$ are projectors on orthogonal  spaces such that  for example $P_A P_B = 0$. 
For a given spin sector  with $d = 2$ the integration over the fermionic degrees of freedom gives \cite{White89,Assaad08_rev}:
\begin{equation}
	\langle \langle \hat{O}  \rangle \rangle_{\pmb{s}^{(1)}, \pmb{s}^{(2)}}    =  \frac{  \det \left[1 +   U^{(2)}_{\pmb{s}^{(2)} } U^{(1)}_{\pmb{s}^{(1)} }  \right] } 	
	             {    \det \left[1 +   U^{(2) }_{\pmb{s}^{(2)}} \right] \det \left[  1 + U^{(1)}_{\pmb{s}^{(1)}}  \right]   2^{-2N_B}  } 
\end{equation}
In the above equation  we have defined 
\begin{equation}
	U^{(r) }_{\pmb{s}^{(r)} }= 
	 \prod_{\tau=1}^{L_{\tau}}  e^{-\Delta \tau T^{(r)}/2}   e^{ V^{(r)} ( \pmb{s}_{\tau}^{(r)}) } e^{-\Delta \tau T^{(r)}/2}.
\end{equation}
Since the equal time Green function \cite{Assaad08_rev} in each replica reads, 
\begin{equation}
      G^{(r) }_{\pmb{s}^{(r)} }  = \left[ 1 + U^{(r) }_{\pmb{s}^{(r)} }\right]^{-1}
\end{equation}
we can  see, after some algebra, that 
\begin{widetext}
\begin{equation}
	\langle \langle \hat{O}  \rangle \rangle_{\pmb{s}^{(1)}, \pmb{s}^{(2)}}   =  
	  \det \left[      P_A  \left( 2 G^{(2)}_{\pmb{s}^{(2)} } P_A G^{(1)}_{\pmb{s}^{(1)} }  - G^{(2)}_{\pmb{s}^{(2)} } -   G^{(1)}_{\pmb{s}^{(1)} }  + 1\right) P_A  + P_B + P_{B'} \right].
 \end{equation} 
 \end{widetext}
Since   $P_A, P_B $ and $P_{B'}$ are orthogonal projectors,   the   above determinant reduces to the determinant of the $N_A \times N_A$ matrix
 $ \det \left[  (G^{(2)}_A -1 )  (G^{(1)}_A - 1 )  + G^{(2)}_A G^{(1)}_A \right]   $     where    $G^{(r)}_A $  corresponds to  the Green function  $ G^{(r)}_{\pmb{s}^{(r)} } $   restricted to 
 Wannier states within $\cal{H}_A$.  The above is nothing but the equation put forward by  Grover  \cite{Grover13}. 
\subsection{The general case.}
To show the equivalence for the n$^{th}$ Renyi  entropy one notes that $ \hat{U}^{(r)}_{\pmb{s}^{(r)}} $ acts non trivially in the Hibert space ${\cal H}_A \otimes {\cal H}^{(r)}_{B}$. Hence, 
\begin{equation}
	\hat{U}^{(r)}_{\pmb{s}^{(r)}}   =    \hat{u}^{(r)}_{\pmb{s}^{(r)}}  \bigotimes_{\stackrel{i = 1 }{i \ne r }}^{n}  \hat{1}_B^{(i)}  
\end{equation} 
The same calculation which leads to Eq. (\ref{Rhon.eq})    gives
\begin{equation}
{\text Tr}_{{\cal H}_{tot}} \left[ \hat{U}^{(n)}_{\pmb{s}^{(n)}} \cdots \hat{U}^{(1)}_{\pmb{s}^{(1)}} \right]   = 
{\text Tr}_{{\cal H}_{A}}   \left[ \tilde{\hat{\rho}}_A (\pmb{s}^{(n)})  \cdots \tilde{\hat{\rho}}_A (\pmb{s}^{(1)})  \right]
\end{equation}
where
\begin{equation}
	\tilde{\hat{\rho}}_A (\pmb{s}^{(r)})  =  {\text Tr}_{ {\cal H}^{(r)}_{B}}  \left[   \hat{u}^{(r)}_{\pmb{s}^{(r)}} \right].
\end{equation}
Using the   relation
\begin{equation}
{\text Tr}_{{\cal H}_{tot}} \left[ \hat{U}^{(r)}_{\pmb{s}^{(r)}} \right]  = {\text Tr}_{{\cal H}_{A}\otimes {\cal H}^{(r)}_{B}} \left[ \hat{u}^{(r)}_{\pmb{s}^{(r)}} \right] d^{(n-1)N_B}
\end{equation}
one obtains: 
\begin{equation}
\langle \langle \hat{O}  \rangle \rangle_{\pmb{s}^{(1)}\cdots \pmb{s}^{(n)}} =    {\text Tr}_{{\cal H}_{A} }\left[ {\hat{\rho}}_A (\pmb{s}^{(n)})  \cdots {\hat{\rho}}_A (\pmb{s}^{(1)})  \right]	
\end{equation}
with 
\begin{equation}
\hat{\rho}_A (\pmb{s}^{(r)})  =   \frac{ {\text Tr}_{ {\cal H}^{(r)}_{B}}  \left[   \hat{u}^{(r)}_{\pmb{s}^{(r)}} \right] }{  {\text Tr}_{ {\cal H}_{A}\otimes {\cal H}^{(r)}_{B}} \left[ \hat{u}^{(r)}_{\pmb{s}^{(r)}} \right]  }.
\end{equation}
$ \hat{\rho}_A (\pmb{s}^{(r)}) $ is an operator acting in ${\cal H}_{A}$. 
For a fixed Hubbard Stratonovitch configuration,  $ \hat{u}^{(r)}_{\pmb{s}^{(r)}} $  is a  single particle propagator such that Wick's theorem applies.  As pointed out in \cite{Grover13} 
it has a  Gaussian representation  uniquely defined by the Green function  $G_A^{(r)} $ given at the end of  the previous sub-section. 
In particular:
\begin{equation}
	\hat{\rho}_A (\pmb{s}^{(r)})   = \det \left( 1 - G_A^{(r)} \right)   e^{ -\hat{\pmb{a}}^{\dagger}  \ln \left(  \frac{1 - G_A^{(r)} } {G_A^{(r)}}   \right)   \hat{\pmb{a}}   } 
 \end{equation}  
 where $ \hat{\pmb{a}} $ is a vector of fermionic annihilation operators running over all single particle states of the Hilbert space $\cal{H}_A $.
Taking the trace over $\cal{H}_{A}$ gives: 
\begin{eqnarray}
\langle \langle \hat{O}  & & \rangle \rangle_{\pmb{s}^{(1)}\cdots \pmb{s}^{(n)}} =    \\ 
& & \prod_{r=1}^{n} \det\left( 1 - G_A^{(r)}  \right)   \det   \left( 1 + \prod_{r=1}	^{n} \frac{G_A^{(r)}}{1 - G_A^{(r)}} \right) \nonumber
\end{eqnarray}
which is nothing but the general result of Ref. \cite{Grover13}.
 \section{Entanglement spectra}  
 \label{Entanglemet.Sec} 
 In  Ref. \cite{Assaad13a}   we proposed to compute the entanglement spectrum by considering the  replica time displaced correlation function: 
\begin{eqnarray}
\label{replica.eq}
   S^{E}_{O}(\tau_\text{E})
      & \equiv & \langle \hat{O}^{\dagger} (\tau_\text{E}) \hat{O} \rangle_\text{A} \nonumber\\
      & \equiv & \frac{{\rm Tr }_{{\cal H}_\text{A}} \left[ \hat{\rho}_{A}^ {(n-\tau_\text{E})}  \hat{O}^{\dagger} \hat{\rho}_A^ {\tau_\text{E} \ } \hat{O} \right]} 
         { {\rm Tr }_{{\cal H}_\text{A}} \left[ \hat{\rho}_A^ {n  } \right] }\;,
\end{eqnarray}
for an operator ${\hat{O} \in {\cal H}_\text{A}}$. Here $\tau_\text{E}$ and $n$ are integers with ${\tau_\text{E} < n}$. 
Within the gaussian approach \cite{Grover13} we can use the representation of  the reduced density matrix, 
\begin{equation}
      \hat{\rho}_A  =  \sum_{\pmb{s}}  P(\pmb{s})  \hat{\rho}_A (\pmb{s}), 
\end{equation}
introduce $n$ replicas  and obtain:
\begin{widetext}
\begin{equation}
	S^{E}_{O}(\tau_\text{E})  = \frac{ \sum_{\pmb{s}^{(1)}  \cdots \pmb{s}^{(n)}  }  P(\pmb{s}^{(1)},\cdots, \pmb{s}^{(n)})  {\text Tr}_ {{\cal H}_{A} } \left[
	           \hat{\rho}_A (\pmb{s}^{(n)}) \cdots  \hat{\rho}_A(\pmb{s}^{(\tau_E +1)}) \hat{O}^{\dagger}  \hat{\rho}_A (\pmb{s}^{(\tau_E)})  \cdots  \hat{\rho}_A (\pmb{s}^{(1)}) \hat{O}     \right] } { \sum_{\pmb{s}^{(1)}  \cdots \pmb{s}^{(n)} } P(\pmb{s}^{(1)},\cdots, \pmb{s}^{(n)}) {\text Tr}_{{\cal H}_{A}}     \left[   \hat{\rho}_A (\pmb{s}^{(n)}) \cdots  \hat{\rho}_A(\pmb{s}^{(1)})  \right]  }.
\end{equation}
\end{widetext}
Sampling over $n$-independent simulations   generates configurations distributed according to $  P(\pmb{s}^{(1)},\cdots, \pmb{s}^{(n)}) $ such that in principle one can compute numerator and denominator   within a single simulation   to provide an estimate of the replica time displaced correlation function.   This approach works at weak coupling but fails  in the strong coupling limit due to fluctuations. Essentially, one is  sampling the wrong distribution, $  P(\pmb{s}^{(1)},\cdots, \pmb{s}^{(n)}) $ and  re-weighting with  the factor   
${\text Tr}_{{\cal H}_{A}} \left[   \hat{\rho}_A (\pmb{s}^{(n)}) \cdots  \hat{\rho}_A(\pmb{s}^{(1)})  \right] $ which accounts for  correlations between the replicas.  Since one can show  that the later quantity is positive  it was proposed in \cite{Assaad13a}  to sample  directly,  $ P(\pmb{s}^{(1)},\cdots, \pmb{s}^{(n)})  {\text Tr}_{{\cal H}_{A}} \left[   \hat{\rho}_A (\pmb{s}^{(n)}) \cdots  \hat{\rho}_A(\pmb{s}^{(1)})  \right] $ so as to access the  strong coupling regime. 

One can achieve this by using the above presented mapping between the gaussian and replica methods. In fact in the extended Hilbert space,  one 
will see that:
\begin{widetext} 
\begin{equation}
	 S^{E}_{O}(\tau_\text{E})    =  \frac{ 
	    \sum_{\pmb{s}^{(1)} \cdots \pmb{s}^{(n)} }  {\text Tr}_{{\cal H}_{tot}} \left[ \hat{U}^{(n)}_{\pmb{s}^{(n)}} \cdots \hat{U}^{(\tau_{\text{E}}+1)}_{\pmb{s}^{(\tau_E +1)}} \hat{O}^{\dagger} \hat{U}^{(\tau_{\text{E}})}_{\pmb{s}^{(\tau_E)}}  \cdots \hat{U}^{(1)}_{\pmb{s}^{(1)}} \hat{O} \right]} 
	             {\sum_{\pmb{s}^{(1)} \cdots \pmb{s}^{(n)} }  {\text Tr}_{{\cal H}_{tot}} \left[ \hat{U}^{(n)}_{\pmb{s}^{(n)}}  \cdots \hat{U}^{(1)}_{\pmb{s}^{(1)}}  \right]}     
\end{equation} 
\end{widetext} 
such that: 
\begin{equation}
 \langle O (\tau_E) O \rangle_{A}    =   \sum_{\pmb{s}^{(1)} \cdots \pmb{s}^{(n)} } \tilde{P}( \pmb{s}^{(1)} \cdots \pmb{s}^{(n)} )     \langle \langle O (\tau_E) O \rangle \rangle_{\pmb{s}^{(1)} \cdots \pmb{s}^{(n)} }  
 \end{equation}
 with 
\begin{equation}
 	\tilde{P}( \pmb{s}^{(1)} \cdots \pmb{s}^{(n)} )   = \frac{ {\text Tr}_{{\cal H}_{tot}}  \left[ \hat{U}^{(n)}_{\pmb{s}^{(n)}}  \cdots \hat{U}^{(1)}_{\pmb{s}^{(1)}}  \right] }
	                            {\sum_{\pmb{s}^{(1)} \cdots \pmb{s}^{(n)} }  {\text Tr}_{{\cal H}_{tot}} \left[ \hat{U}^{(n)}_{\pmb{s}^{(n)}}  \cdots \hat{U}^{(1)}_{\pmb{s}^{(1)}}  \right]}
\end{equation}
 and 
\begin{eqnarray}
	 \langle \langle O (\tau_E) O & &   \rangle \rangle_{\pmb{s}^{(1)} \cdots \pmb{s}^{(n)} }   =  \\
	 & &     \frac{ {\text Tr}_{{\cal H}_{tot}} \left[ \hat{U}^{(n)}_{\pmb{s}^{(n)}} \cdots \hat{U}^{(\tau_{\text{E}}+1)}_{\pmb{s}^{(\tau_E +1)}} \hat{O}^{\dagger} \hat{U}^{(\tau_{\text{E}})}_{\pmb{s}^{(\tau_E)}}  \cdots \hat{U}^{(1)}_{\pmb{s}^{(1)}} \hat{O} \right]} { {\text Tr}_{{\cal H}_{tot}} \left[ \hat{U}^{(n)}_{\pmb{s}^{(n)}}  \cdots \hat{U}^{(1)}_{\pmb{s}^{(1)}}  \right] }.
	    \nonumber
\end{eqnarray}
The above corresponds to a {\it standard}  calculation of an imaginary  time displaced correlation function in the extended Hilbert space  at temperature $n\beta$ albeit  with 
an  imaginary time  dependent  Hamiltonian.    This quantity can readily be implemented in standard auxiliary field  finite temperature quantum Monte-Carlo methods.  The above formulation is however not restricted to fermions.   In  fact,  it  carries over  to  bosonic systems amenable to  stochastic simulations   within, for example,  the  stochastic series expansion  algorithm \cite{Sandvik02}.  

\section{Results}

\begin{figure}
   \includegraphics[width=0.9\linewidth]{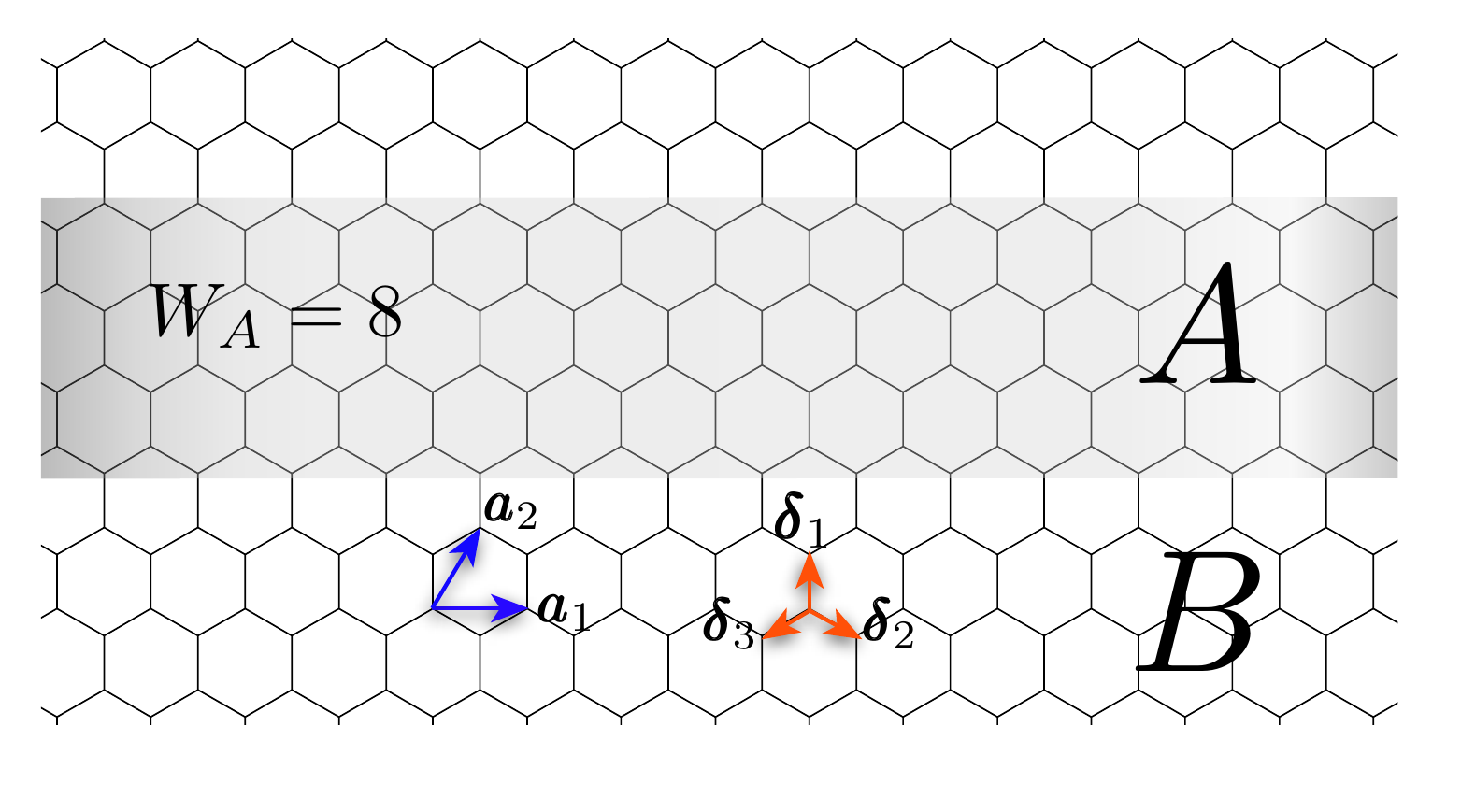} 
   \caption{  Honeycomb lattice.  For an $L \times L$ lattice, we consider  periodic boundaries: 
  $  \pmb{c}_{\pmb{i} + L \pmb{a}_1} =  \pmb{c}_{\pmb{i}}$ and $  \pmb{c}_{\pmb{i} + L \pmb{a}_2} =  \pmb{c}_{\pmb{i}}$.  The real space partitioning  breaks translation symmetry in the $\pmb{a}_2$ direction  but not along $\pmb{a}_1$.
   \label{Latt.fig}}
\end{figure}

\begin{figure}
   \includegraphics[width=0.9\linewidth]{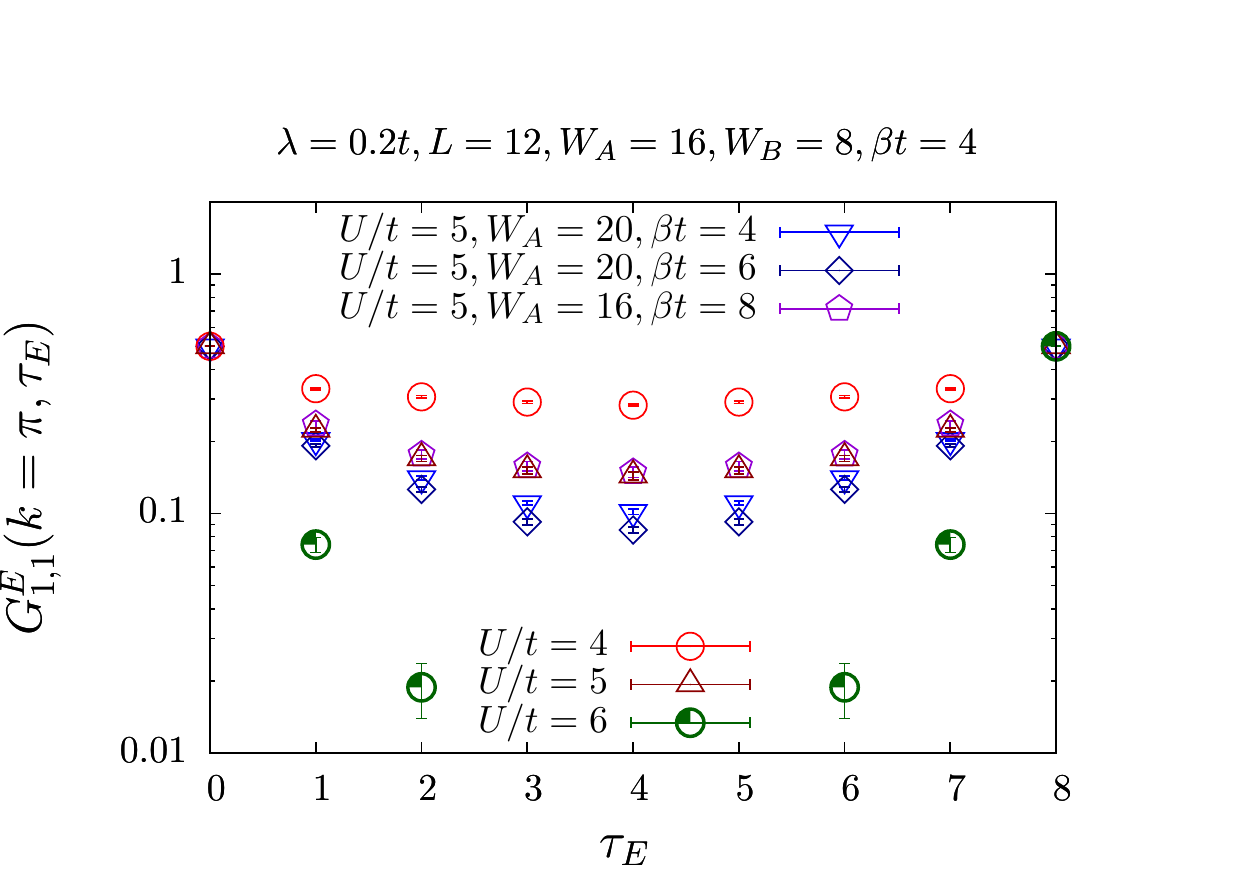} 
   \caption{Entanglement single particle Green function. Here we concentrate on the time  reversal symmetric momentum $k = \pi$ and orbital $m=1$ lying  on the boundary of subsystem A. 
   \label{Green.fig}}
\end{figure}
To illustrate the fact that we are able to   access the strong coupling regime, we  consider the interaction driven quantum phase transition in the Kane-Mele Hubbard model.   The model is defined on the Honeycomb lattice.   Using the spinor notation ${\hat{{\mathbf{c}}}^{\dagger}_{i} = \big(\hat{c}^{\dagger}_{\pmb{i}\uparrow}, \hat{c}^{\dagger}_{{\pmb i}\downarrow}\big)}$ it reads 
\begin{align}
\label{eq:KMU}
  \hat{H}_{KMU} =
   \sum_{ {\pmb i},{\pmb j} } \hat{c}^{\dagger}_{\pmb{i}} \left[ t_{ \pmb{i}\pmb{j} } + {\rm i}\, \boldsymbol{\lambda}_{{\pmb i}{\pmb j }} \cdot \boldsymbol{\sigma} \right] \hat{c}^{\phantom{\dag}}_{\pmb j}    +  \frac{U}{2}\sum_{\pmb{i}} \left(  \hat{{\mathbf{c}}}^{\dagger}_{i}  \hat{{\mathbf{c}}}^{\phantom{\dagger}}_{i}   - 1\right)^2 \;.
 \end{align}
The hopping matrix takes non-vanishing values, $-t$, between nearest neighbors of the honeycomb lattice, ${{\pmb i } - {\pmb j} = \pm{\pmb \delta}_1, \pm{\pmb \delta}_2, \pm{\pmb \delta}_3}$ (see Fig.~\ref{Latt.fig}) and the  intrinsic spin-orbit term is given by
\begin{equation}
	{\pmb \lambda}_{ \pmb{i}\pmb{j} } = \lambda
   \left\{
      \begin{array}{cl}
         \frac{ (\pmb{ i} - \pmb{r}) \times (\pmb{ r} - \pmb{j}) }{ \left|(\pmb{ i} - \pmb{r}) \times (\pmb{ r} - \pmb{j}) \right| } & \text{if } \pmb{i},\pmb{j} \text{ are n.n.n.} \\
         0 & \text{otherwise} 
      \end{array}\;,
   \right.
\end{equation} 
where ${\pmb r} $ is the intermediate site involved in the next nearest neighbor (n.n.n.) hopping process from site $ \pmb{i} $ to $\pmb{j}$.  
At  $\lambda = 0.2 t$  the model shows a zero temperature  phase transition  between  a quantum spin Hall state and a quantum antiferromagnetic at $U_c/t =  5.71(2)$ \cite{Toldin14}.  The  quantum phase transition is well understood and belongs to the 3D XY universality class.  Here, we show that we can detect this  phase transition in the entanglement spectrum.  In the absence of interactions the entanglement Hamiltonian is adiabatically linked to the  original one such that both have the same topological properties \cite{Fidkowski10,Turner10}.  Thereby    the entanglement  Hamiltonian corresponding to a real space partitioning of the system  should show edge states.     

\begin{figure}
\includegraphics[width=\linewidth]{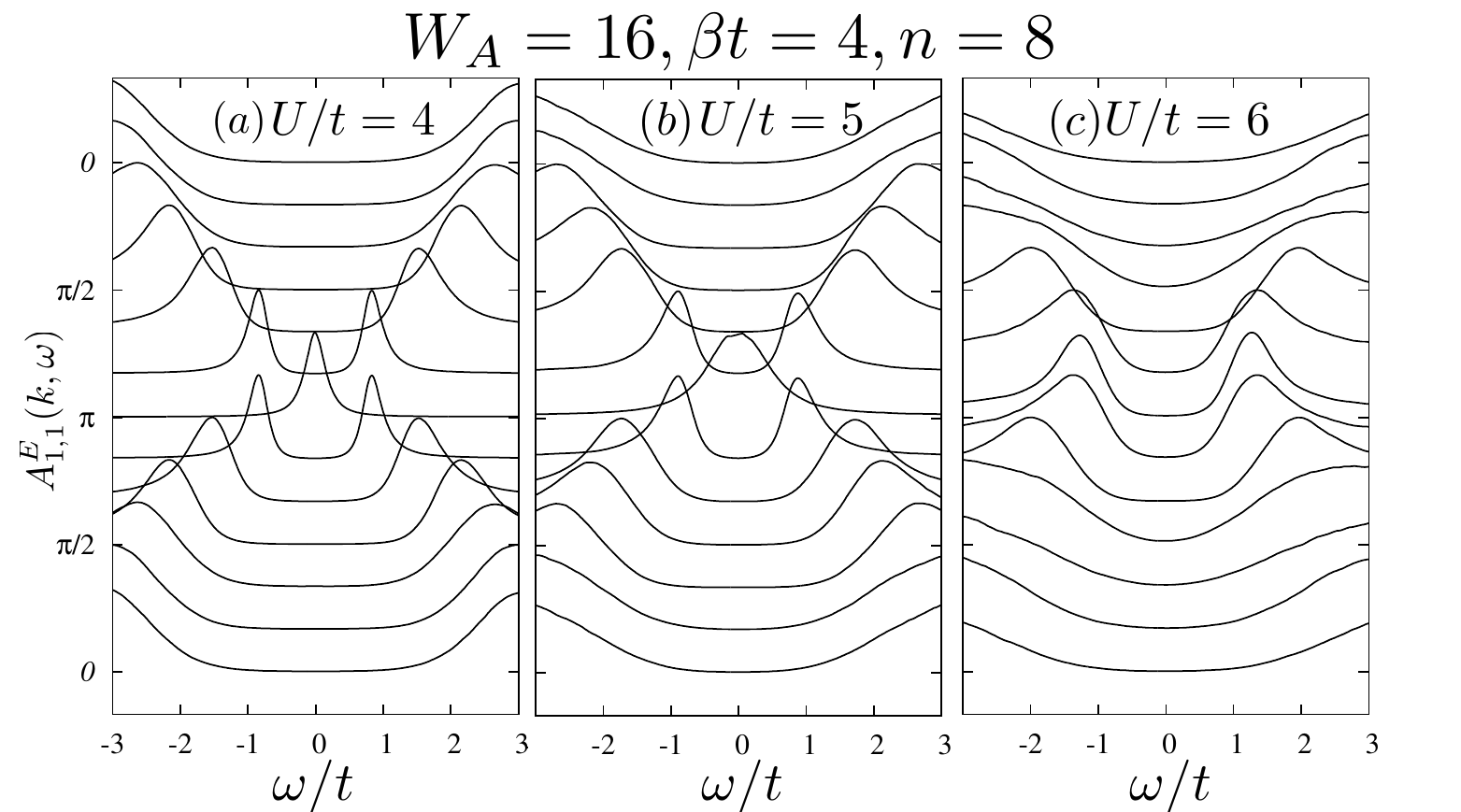} 
   \caption{Entanglement spectral function as a function of   the Hubbard $U$.  For each $k$-point the sum rule $\int \text{d}  \omega A _{1,1}^{E}(k,\omega) = 1 $ holds.  In the plot, we have normalized the peak  hight to unity. 
   \label{Akom.fig}}
\end{figure}

\begin{figure}
\includegraphics[width=\linewidth]{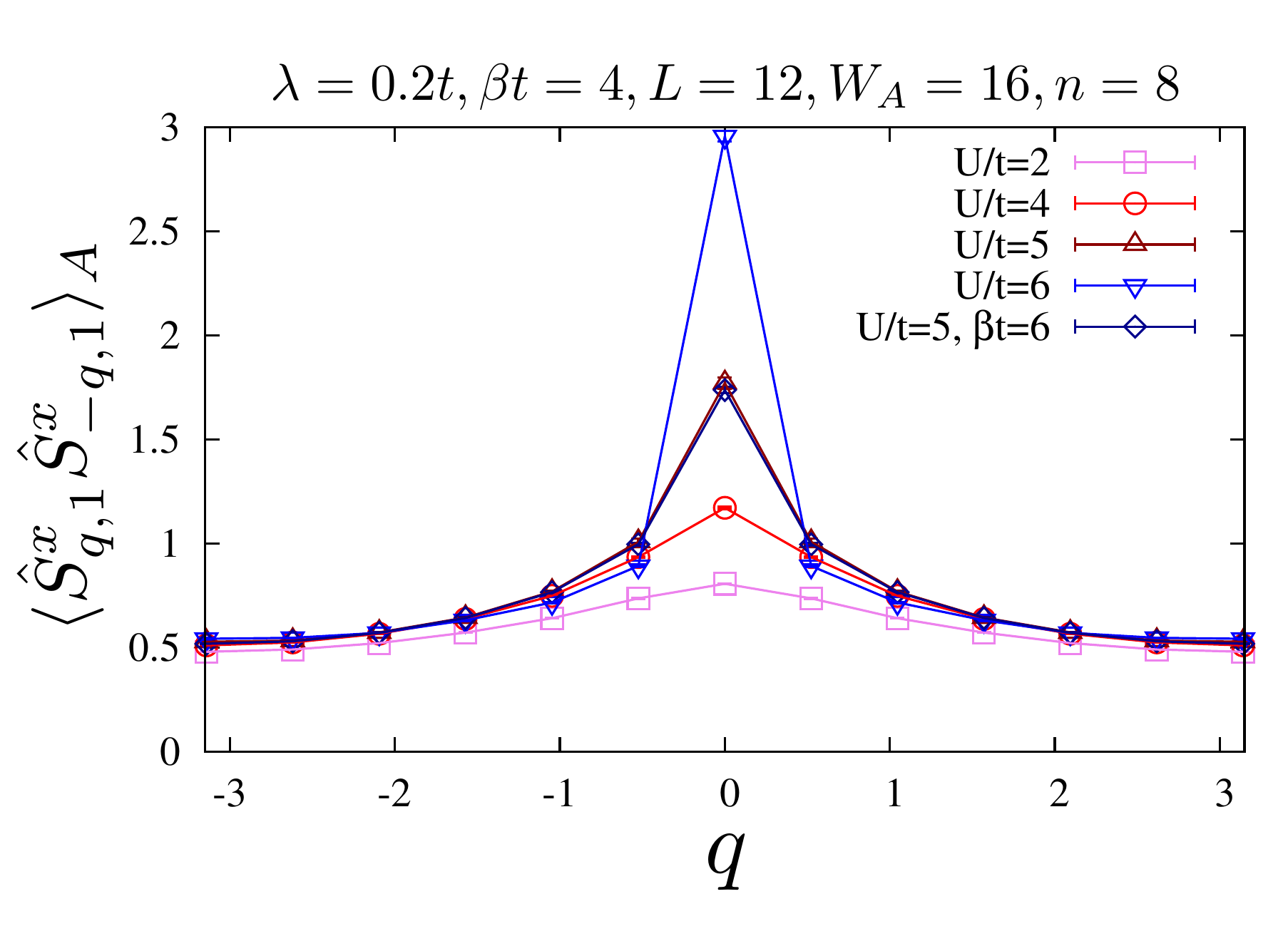} 
   \caption{  x-component of the spin-spin correlation function  taken on the edge of subsystem A corresponding to orbital index $m=1$. 
   \label{Spineq.fig}}
\end{figure}

Fig. \ref{Green.fig}   shows the  single particle entanglement  replica time displaced  Green function, 
\begin{equation}
	G^{E}_{m,m'}(k,\tau_E) = \frac{1}{2}\sum_{\sigma} \langle  \hat{a}_{k,m,\sigma}^{\dagger}(\tau_E)  \hat{a}_{k,m',\sigma} \rangle_A.
\end{equation}
The real space cut we consider is translationally  invariant in the $\pmb{a}_1$ lattice direction. Thereby, $k=\pmb{k} \cdot \pmb{a}_1$ is a good quantum number which we can use to classify the data. The label $m$ is an {\it orbital} index  running across  the width, $W_A$, of the cut.    In Fig.~\ref{Green.fig}   we consider a $12 \times 12 $ lattice   with $n=8$ replicas, $W_A =16,20$ and  inverse temperatures $\beta t = 4,6,8 $.  Note that $W_A + W_B = 2L$ such that at $n=8$ our largest simulations  have $960$ sites at an effective inverse temperature $n\beta t = 64$.  All our simulations are carried out at a finite imaginary time step $\Delta \tau t = 0.1$.   In Fig.~\ref{Green.fig}  we concentrate on the time reversal symmetric momentum $k = \pi$  and orbital corresponding to the edge of the cut, $m=1$. Since  particle hole symmetry  is present  in the model,  the Dirac cone  is pinned at the fermi energy. Thereby  a  signature of the topological phase, is  a non-decaying single particle entanglement Green function as a function of the replica time $\tau_E$.   At $U/t=5$  we have considered  various temperatures and values of $W_A$.  As apparent, as a function of increasing  $W_A$ and thereby decreasing $W_B$,  $G^{E}_{1,1}(\pi,\tau_E) $  decays more quickly.  This may be assigned to  edge-edge correlations across  the $B$ subsystem.    The phase transition is triggered by the onset of magnetic correlations which at $T=0$ develop long range order  beyond $U_c$ thereby breaking time reversal symmetry.    As a consequence, enhancing the temperature will reduced  the magnetic correlation length,   stabilize the topological state and show a less pronounced decay in $G^{E}_{1,1}(\pi,\tau_E) $.  

To  obtain a better overview of the data, we can  define an entanglement spectral function by analytical continuation of the replica time data: 
\begin{equation}
\label{Analytical_continuation.eq}
   G^{E}_{m,m}(k,\tau_E)   = \frac{1}{\pi} \int {\rm d} \omega  \frac{{\rm e}^{- \tau_\text{E} \omega} }{ 1 + {\rm e}^{-\tau_\text{E}  \omega} } A^{E}_{m,m}(k,\omega). 
\end{equation}
To carry out this step, we have used the stochastic Maximum Entropy approach \cite{Sandvik98,Beach04a}.  Our results are plotted in Fig.~\ref{Akom.fig}. 
As apparent  below    $U_c/t = 5.71(2)$  we observe a single Dirac cone  and beyond the phase transition   a gap in the entanglement spectrum opens.

The gap in the entanglement spectral function  stems from the onset of spin-spin correlations. The equal  time spin-spin correlations of the entanglement Hamiltonian can be computed from 
\begin{equation}
	\langle \hat{S}^{x}_{q,m} \hat{S}^{x}_{-q,m'}  \rangle_A \equiv   \frac{ \text {Tr} \left[ \hat{\rho}_A^n \hat{S}^{x}_{q,m} \hat{S}^{x}_{-q,m'}   \right] } {  \text {Tr} \left[ \hat{\rho}_A^n    \right]}
\end{equation}
with $\hat{S}^{x}_{q,m}  = \frac{1}{2\sqrt{L}}   \sum_{i_x} e^{i q i_x} \hat{\pmb{a}}^{\dagger}_{(i_x,m) } \sigma_x \hat{\pmb{a}}_{(i_x,m) }$ corresponding to a partial  Fourier transformation of the x-component of the spin-operator.  
As apparent from Fig.~\ref{Spineq.fig}   a sharp peak at  $q = \pi$ emerges  beyond the transition at $U_c/t=5.71(2)$
\section{Conclusion}
\label{Conc.Sec}

In this article we have shown that the  two methods put forward to compute   the n$^{th}$  Renyi entropies in  determinant QMC  methods for fermions are  in essence identical. Starting with the replica scheme proposed in \cite{Hastings10,Humeniuk12}  and adapted to determinant  \cite{Broecker14} and continuous time  \cite{Wang14}  QMC, we can derive the free fermion or  gaussian approach put forward by Grover \cite{Grover13}.   The two methods differ in the sampling strategy. The  gaussian approach samples $n$ independent replicas  and correlations between the replicas are taken into account by  re-weighting. The swap algorithm  formulates the QMC in an extended Hilbert space   thereby explicitly sampling  correlations between replicas.  The mapping between both methods shows how to formulate  numerical simulations to access entanglement spectra at strong coupling by carrying out a standard  simulation within the extended  Hilbert space  of subsystem  $A$ and n replicas of subsystem $B$  albeit with a time dependent Hamiltonian. In contrast to  our former approach described in \cite{Assaad13a}  the present formulation does not suffer from uncontrollable fluctuations in the strong coupling regime. We  were able to study  aspects of the entanglement spectrum in the  correlation driven quantum phase transition between a topological insulator  and quantum antiferromagnetic    as realized in the Kane-Mele Hubbard model.
	 The present  formulation is  numerically expensive since    the total number of sites scales as $N_A + n N_B$ where $n$   corresponds to the number of replicas and $N_A$  ($N_B$)  the number of sites in  subsystems  $A$  ($B$).    The  structure of the imaginary time evolution allows for many optimization strategies. Nevertheless, the overall computational effort  scales as $ n \beta  \left(N_A + n N_B \right)^3$.   Our approach to compute the  entanglement spectrum is not  specific to simulations of  fermonic systems in the realm of  determinant QMC methods.  In fact it can be adapted to bosonic  systems within, for example, the SSE \cite{Sandvik02} approach. Since these methods have a very favorable scaling, $    n \beta  \left(N_A + n N_B \right) $,  introducing many replicas is not as expensive as for fermions.

I would like to thank T. Lang and F. Parisen Toldin  for many invaluable  discussions,  and P. Br\"ocker, T. Grover and Lei Wang for comments.  We thank the LRZ-M\"unich and the J\"ulich Supercomputing center for generous allocation of CPU time. Financial support from the DFG grant AS120/9-1 is acknowledged.

 \end{document}